\documentclass{article}

\usepackage{standalone}
\usepackage{etoolbox}

\usepackage{microtype}

\usepackage{natbib}

\usepackage{graphicx}
\usepackage{subcaption}
\usepackage[table]{xcolor}

\usepackage{chngcntr}

\usepackage{booktabs} 

\makeatletter
\@ifundefined{c@rownum}{%
  \let\c@rownum\rownum
}{}
\@ifundefined{therownum}{%
  \def\therownum{\@arabic\rownum}%
}{}
\makeatother

\usepackage{siunitx}
\usepackage{xspace}
\makeatletter
\DeclareRobustCommand\onedot{\futurelet\@let@token\@onedot}
\def\@onedot{\ifx\@let@token.\else.\null\fi\xspace}

\def\eg{e.g\onedot} 
\def\ie{i.e\onedot}

\makeatother

\usepackage{hyperref}

\usepackage[accepted]{icml2019}

\makeatletter
\patchcmd{\@makecaption}
{\vskip 10pt}
{\vskip 2.5pt plus 5pt}
{}{}
\makeatother

\hypersetup{
  colorlinks,
  linkcolor = red!50!black,
  citecolor = mydarkblue,
  urlcolor  = magenta!75!black,
}

\icmltitlerunning{A Halo Merger Tree Generation and Evaluation Framework}

\newcommand{\Msun}{M_\odot}
\newcommand{\nvar}{\mathrm{n_{var}}}

\begin{document}

\twocolumn[%
\icmltitle{A Halo Merger Tree Generation and Evaluation Framework}



\icmlsetsymbol{equal}{*}

\begin{icmlauthorlist}
\icmlauthor{Sandra Robles}{uam,mel}
\icmlauthor{Jonathan S. G\'omez}{puc}
\icmlauthor{Ad\'in Ram\'irez~Rivera}{ic}
\icmlauthor{Jenny A. Gonz\'alez}{puc}
\icmlauthor{Nelson D. Padilla}{puc}
\icmlauthor{Diego Dujovne}{udp}
\end{icmlauthorlist}

\icmlaffiliation{ic}{Institute of Computing, University of Campinas, S\~ao Paulo, Brazil.}
\icmlaffiliation{uam}{Departamento de F\'isica Te\'orica, Universidad Aut\'onoma de Madrid, Madrid, Spain.}
\icmlaffiliation{mel}{ARC Centre of Excellence for Particle Physics at the Terascale, School of Physics, The University of Melbourne, Victoria, Australia.}
\icmlaffiliation{puc}{Instituto de Astrof\'isica, Pontificia Universidad Cat\'olica de Chile, Santiago, Chile.}
\icmlaffiliation{udp}{Escuela de Inform\'atica y Telecomunicaciones, Universidad Diego Portales, Santiago, Chile}

\icmlcorrespondingauthor{Sandra Robles}{sandra.robles@unimelb.edu.au}
\icmlcorrespondingauthor{Ad\'in Ram\'irez~Rivera}{adin@ic.unicamp.br}

\icmlkeywords{Halo Merger Tree Generation, Generative Adversarial Networks}

\vskip 0.3in
]



\printAffiliationsAndNotice{}  

\begin{abstract}
Semi-analytic models are best suited to compare galaxy formation and evolution theories with observations. 
These models rely heavily on halo merger trees, and their realistic features (\ie, no drastic changes on halo mass or jumps on physical locations). 
Our aim is to provide a new framework for halo merger tree generation that takes advantage of the results of large volume simulations, with a modest computational cost.
We treat halo merger tree construction as a matrix generation problem, and propose a Generative Adversarial Network that learns to generate realistic halo merger trees. 
We evaluate our proposal on merger trees from the EAGLE simulation suite, and show the quality of the generated trees. 
\end{abstract}

\section{Introduction}
\label{sec:intro}

In cosmology and astrophysics, galaxy formation and evolution is a complex non-linear problem that entails very different physical phenomena spanning a wide range of scales, from large scale structure formation, dark matter (DM) gravitational collapse in halos down to  star formation, and metal enrichment among others.   
Two strategies are used to tackle this problem, semi-analytic models (SAMs) \cite{White1991} and hydrodynamical simulations \cite{Carlberg1990,Katz1992}. 

Large-scale hydrodynamical simulations solve the equations of gravity and fluid dynamics simultaneously. 
Nevertheless, high-resolution simulations are extremely computational resource intensive. SAMs, on the other hand, describe physical processes behind galaxy formation and evolution in a simpler way and require less CPU time. 
Furthermore, they can simulate much larger cosmological volumes than hydrodynamical simulations and consequently they can be tuned to  observations.
These models can be populated with halo merger trees and then phenomenological prescriptions of the baryonic physics governing galaxy formation are introduced. 

A halo merger tree describes the hierarchical formation history of a DM halo by tracing all its progenitors back in time. 
In the standard model of cosmology, large DM halos are formed by the coalescence of smaller progenitors. 
Therefore, halo merger trees encapsulate the growth and merger history of DM halos. 
If the progenitors contain central galaxies, halo mergers\footnote{We refer to them as `mergers' throughout this paper.  In that sense, we are describing the event of halos merging together.} eventually give rise to galaxy mergers. 
In this way, galaxy evolution is directly linked to the halo merger history.

Two different methodologies are used to produce halo merger trees.  
The first method takes advantage of the extended Press-Schechter formalism \cite{Bond1991} and Monte Carlo simulations. It allows rapid  merger tree construction in large volumes with high mass resolutions  \cite{Somerville1999,Cole2000,Somerville2008,Benson2010,Ricciardelli2010}, but a single tree per halo can be  built at a time. 
In addition, this method often yields  mass assembly histories that are  in disagreement with simulations  \cite{Jiang2014J}.

Despite being computationally intensive, the most popular method for merger tree construction is based on high-resolution DM only ($N$-body) simulations, since it results in a more realistic evolutionary history of the halos and yields thousands of merger trees at once  \cite{Kauffmann1999,Hatton2003,DeLucia2004,Croton2006,Bower2006,Guo2011,Lee2014}. 
An important caveat of merger tree construction using simulations is the limitation in  mass resolution. 
Since massive halos are better resolved than those with low mass, the former can be traced down to progenitor masses that are a smaller fraction of the final halo mass than the latter.  Merger trees are built in two steps. First,  each  output time-step of the simulation (snapshot) is scanned to find DM collapsed structures, halos and produce halo catalogs. Then, or simultaneously to the DM halo search, halos are linked across different snapshots giving us the merger history of each halo (the structure known as merger tree), where  the last leaf  (last descendent) of the main branch (the longest branch) is the halo whose assembly history the tree encloses.
The remaining leaves of the tree are the progenitors of the aforementioned halo.

Halo merger trees play an important role in modern galaxy formation theory, and are the key ingredient of SAMs. 
More specifically, mock halo catalogs from one step in time (snapshot) of a DM only cosmological simulation are the usual inputs of SAMs. Next, these DM halos are populated with galaxies using a given baryonic physics prescription. A self-consistent evolution of these galaxies requires the knowledge of the growth and mass assembly history of the DM halo that hosts them, which is precisely what halo merger trees provide. 
In this sense, SAMs rely heavily on both a precise halo identification and, realistic and well-constructed merger trees, with no drastic changes in the halo mass, or jumps in physical location from one snapshot to the next.

Nowadays, SAMs are best suited to compare theoretical predictions with galaxy surveys, thanks to their flexibility to explore physical phenomena in different ways with relatively computational inexpensive runs \citep[see \eg][]{Cole1991,Lacey1991,White1991,Kauffmann1993a,Kauffmann1993b,Kauffmann1999,Cole1994,Cole2000,Bower2006,Somerville2008,Lagos2018}. 
As mentioned before, the standard method to produce realistic merger trees is based on high resolution $N$-body simulations that require long runs. 
Our main contribution is to provide astrophysicists with a new framework for halo merger tree construction, taking advantage of the best features of large volume simulations, but with a modest computational expense. 
To this end, we designed a Generative Adversarial Network (GAN) \cite{Goodfellow2014} that generates merger trees (represented as matrices) with characteristics of a subset of trees from the EAGLE simulation suite \cite{Shaye2015,Crain2015}. 
Additionally, we introduce a set of criteria to assess the quality of the generated merger trees when compared to those obtained from simulations.

\section{Halo Merger Tree Generation}

\begin{figure}[t!]
\centering
\includegraphics[width=0.975\linewidth]{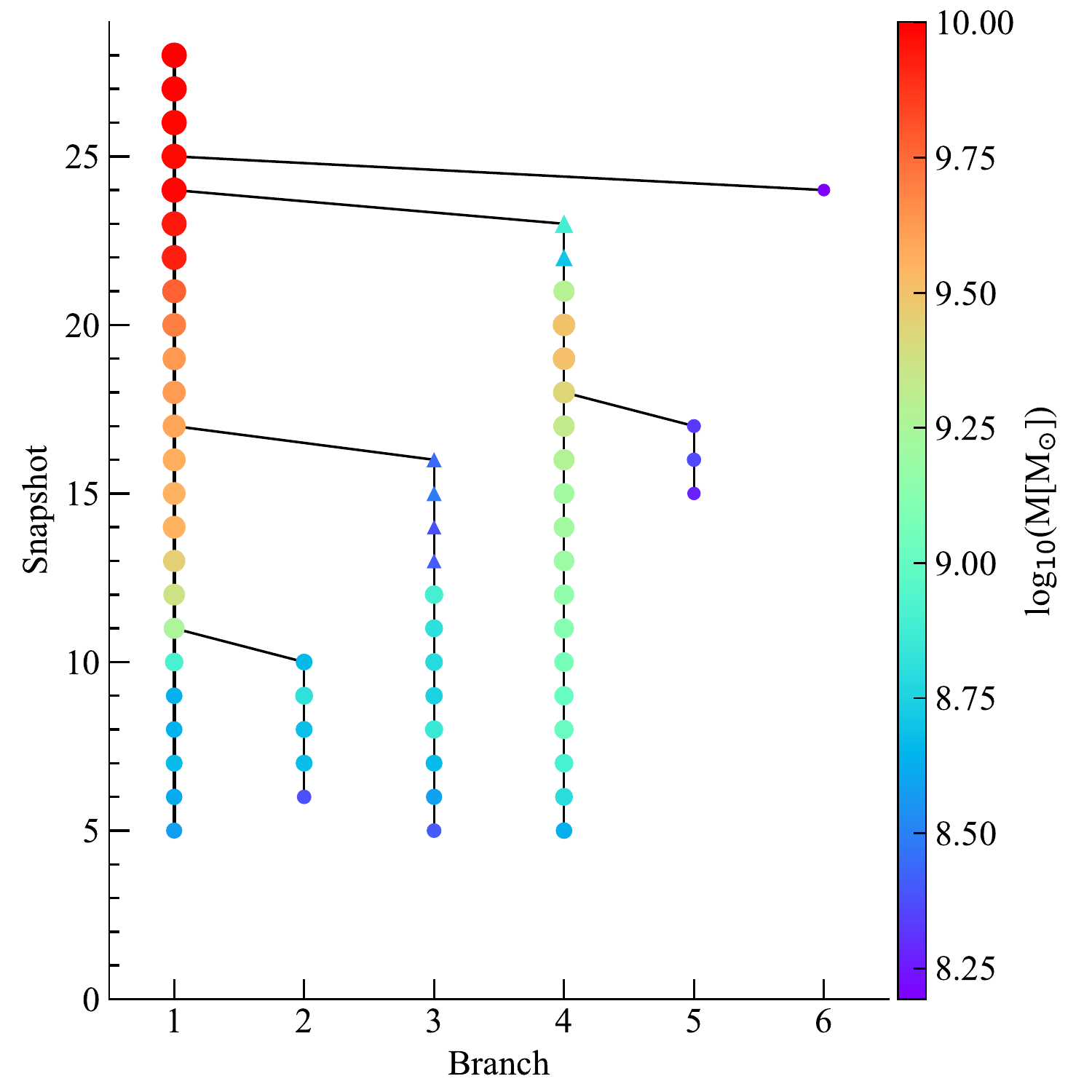}
\includegraphics[width=0.975\linewidth]{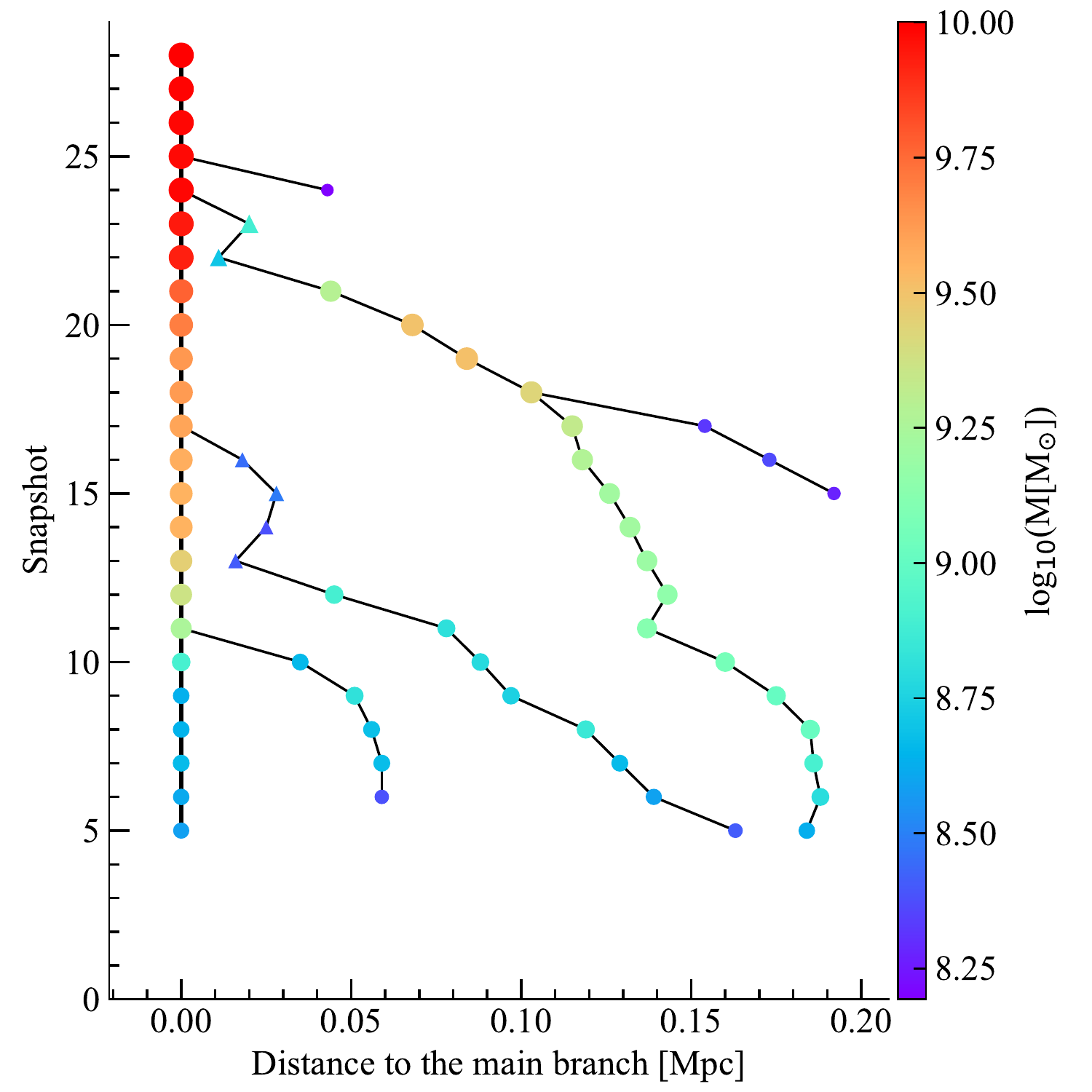}
\caption{%
  Top: Generated merger tree, the color map denotes progenitor masses and the progenitor type is represented by  circles (main halos) and triangles (subhalos).
  Bottom: Same merger tree  in the plane snapshot vs.\ distance.
} 
\label{fig:mergertree}
\end{figure}

\begin{figure*}[tb]
\centering
\vspace{10pt}
\includegraphics[width=0.8\linewidth]{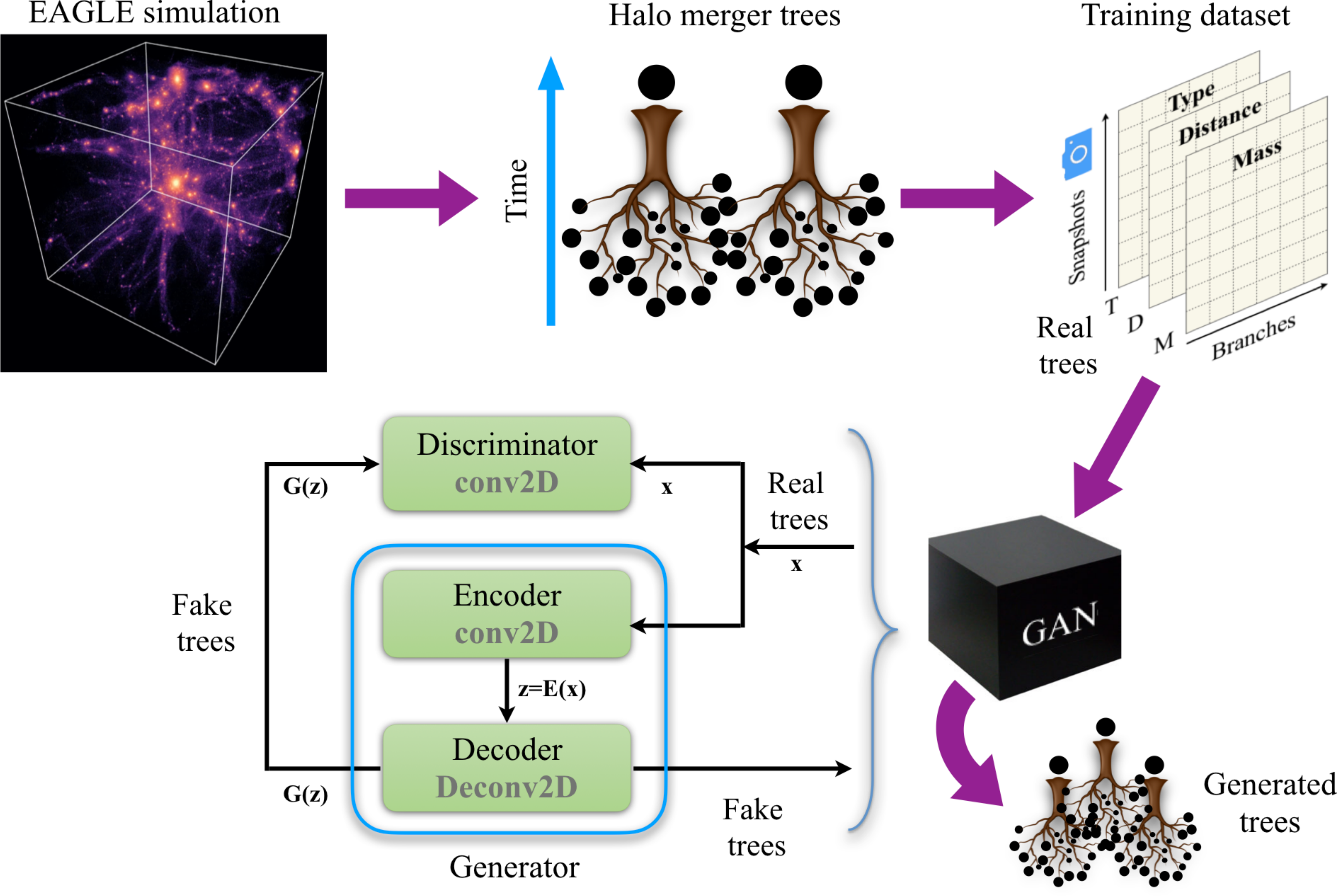}
\caption{%
Halo merger tree generation process. 
} 
\label{fig:mt_generation}
\end{figure*}

Halo merger trees are the backbone of SAMs. 
With the upcoming of new major observational facilities, robust predictions from SAMs will be needed to understand the nature of the processes that could imprint observational features on galaxies. 
Hence, our goal is to learn a lean generative model, based on GANs, that generates merger trees closer to the training data. 
To that end, first, we represent merger trees as matrices. 
Next, we propose a learning framework to train our model with different sets of variables.

\textbf{Merger Tree Representation.}
We selected halo merger trees for central subhalos at $z=0$ (current time of Universe) with masses in the range $10^{10}\Msun \leq M \leq 10^{11}\Msun$ from the EAGLE reference simulation with comoving cubic box length of \SI{100}{Mpc} and $1504^3$ particles. 
This simulation supplies a large enough amount of trees for training purposes. 
We obtained the trees by traversing the subhalo table of the EAGLE public catalog \cite{McAlpine2016}. 
The EAGLE simulation suite is a set of cosmological hydrodynamical simulations with cosmological parameters taken from \citet{Planck2014}, that track the evolution of both, dark matter and baryonic particles, from redshift $z=127$ to $z=0$. 

Next, we stored each relevant variable for consistent merger tree description, namely mass of the halo progenitors, distance of every progenitor in a branch to the corresponding progenitor in the main branch and a discrete variable that indicates if the progenitor is a main halo or a subhalo,\footnote{Note, that halos can be nested, \ie, they can contain substructures. The halo at the center of the gravitational potential is referred to as main or central halo. The others are called  subhalos, often regarded as orbiting main halos.} in a matrix format that reflects the tree structure (see Fig.~\ref{fig:mergertree}), one matrix per each variable. 
Columns in this format represent the branches of a tree, and rows correspond to $29$~snapshots between redshifts $z=20$ (snapshot~$0$) and $z=0$ (snapshot~$28$, present day). 
The first column is the so-called main branch that is the longest branch of a tree. 
At snapshot~$28$, this branch contains the last descendant, \ie, the halo whose mass assembly history the merger tree tells about.  
Note that the main branch not necessarily starts at snapshot~$0$, indeed the location in time of the first progenitor depends on the mass of the final halo and the resolution of the simulation. 
Recall that massive halos can be traced down to low mass progenitors at earlier steps in time and hence their merger trees tend to have longer main branches than those of low massive halos. 
Progenitors in other branches eventually merge with the corresponding progenitor in the main branch (same snapshot) in the next time step. 
Matrix elements are filled following the structure of a tree.
Zeros in each matrix denote no progenitor in that particular branch and snapshot. 

\textbf{Neural Network Model.}
To generate merger trees, we selected a deep convolutional GAN architecture, based on DCGAN \cite{Dosovitskiy2015, Radford2015}. 
More specifically, the generator features an encoder-decoder architecture \cite{Bengio2013} that learns to produce the matrix representation of the halo merger trees.
We detail the architecture in Appendix~\ref{sec:architecture}. 
Our design for this model uses row- and column-wise convolutional filters.
These filters intend to reproduce the operations within a branch (column-wise filters), and between the branches to produce mergers (row-wise filters).

We trained this model with the dataset in matrix form, where the merger tree variables (mass, distance to the main branch, and progenitor type) correspond to input channels. 
The reconstruction loss function that drives the learning process comprises cross-entropy losses applied individually to each of the variables. 
We have restricted our initial dataset for training purposes to trees with only six branches, since they are the most abundant in the aforementioned mass range, and the reconstruction of trees with a fixed number of branches is a good test of the viability of GANs to learn merger tree representations. 
In Fig.~\ref{fig:mt_generation}, we show a diagram that summarizes the entire merger tree generation process. 

\section{Evaluation of Merger Trees}
\label{sec:evaluation}

As stated before, SAMs require well-constructed merger trees in order to make consistent theoretical predictions. Therefore, we developed a set of criteria to evaluate both the quality of the generated trees, and the effect of adding variables or inputs to the GAN on the generated trees. 

First of all, GAN outputs should resemble a real merger tree, \ie, the learned representation should contain no reappearing halos after a merger, the absence of a progenitor in a branch must be represented by a zero value, the last row (snapshot~$28$) should contain only one halo, the last descendant. 
This criterion holds for all the variables considered in this work. 

Merger trees tell the mass assembly history of halos. 
Hence, the first variable to reproduce is the progenitor mass. 
The second criterion takes into account a fair reproduction of the  masses, \ie, masses in a branch should be within the mass range of the dataset used for training.  
It is worth mentioning that progenitor masses in a branch are usually expected to grow monotonically 
or at least preserve its mass,  which is an assumption for SAMs \citep{Lacey1991, White1991, Cole1994, Cole2000, Gonzalez-Perez2014, Lacey2016, Lagos2018}. 
However, there are other variables which can alter that expected behavior, such as the distance to the main branch and the progenitor type. 

The second variable considered is the distance between the centers of a progenitor in a given branch with respect to the progenitor in the main branch at the same step in time. 
In general, this distance should decrease as the merging event approaches, but the exact step in time when this should occur varies with the merger tree. 
Nevertheless, we can extract statistic distributions about the behavior of this input variable for several snapshots before the merger occurs, for both samples the real (training dataset) and generated merger trees (see Appendix~\ref{sec:distance}). 
A way to evaluate whether or not the distances are well reproduced by the learned representation, \ie, there are no dramatic jumps in location, is by comparing the real and fake distributions with the Kolmogorov-Smirnov (KS) test.  
Unlike the mass, there is no criterion for a correct range in which the prediction of the distance between two merging progenitors should lie. 
Note that, together, mass and distance are a measure of the gravitational pull between merging progenitors. 

The third variable accounts for the progenitor type, \ie, a progenitor in a branch can be either a main (central) halo or a subhalo. 
Progenitors in the main branch are expected to be central halos, with some exceptions at earlier snapshots. 
On the other hand, progenitors in other branches can be always main halos or become subhalos a few snapshots before the merger as a consequence of gravitational infall.
Therefore, the fusion is strongly linked to the condition of the progenitors involved being central or subhalos \citep{Diemand2006, Muldrew2011, Han2012, Onions2012, Elahi2011, Onions2013}.  
Once again, the exact step in time this should happen depends on the specific tree. 
The condition of a progenitor being a main or a sub-halo affects the behavior of the other two variables. 
Subhalos are expected to be closer to the progenitor in the main branch they are going to merge, than central halos.  
Hence, subhalo masses are allowed to diminish in time as long they approach the merging event. 
We show an example of the interplay among the three variables in Fig.~\ref{fig:mergertree}, where progenitors become subhalos before the merger in the second and third branches and their masses decrease. 
In fact, in both cases the progenitor mass diminishes one snapshot before they become subhalos, as they approach the progenitor in the main branch (see bottom panel of Fig.~\ref{fig:mergertree}). 
Therefore, the mass gain and loss in a generated tree cannot be penalized arbitrarily in a particular tree, we should compare sample distributions instead. 
More examples of merger trees generated with 3 variables can be found in Appendix~ \ref{sec:examples}. 

In light of the foregoing, a final test to probe that the progenitor type is well generated by our GAN framework is to perform a KS test on the distribution of the number of snapshots a progenitor spend as a subhalo before merging with another progenitor. 

\section{Experiments and Results}
\label{sec:results}

\begin{table}[tb]
\caption{%
  Kolmogorov-Smirnov test for mass, distance to the main branch and number of snapshots that a progenitor spend as a subhalo before the merging event for merger trees generated with 1, 2 and 3~variables. Dashes denote unavailable tests.
}
\label{tab:KStest}
\vspace{0.15in}
\small
\centering
\sisetup{
  table-format=1.2,
}
\begin{tabular}{lSSS}
\toprule
KS Test          & {1 Variable} & {2 Variables} & {3 Variables} \\
\midrule
Mass             & 0.43         & 0.57          & 0.21          \\
MH mass          & {--}         & {--}          & 0.14          \\
SH mass          & {--}         & {--}          & 0.05          \\
\midrule
Distance        & {--}         & 0.05          & 0.06          \\
\midrule
N. snaps. as subhalo   & {--}         & {--}          & 0.04         \\
\bottomrule
\end{tabular}
\end{table}

\begin{figure}[tb]
  \centering
  \includegraphics[width=\linewidth]{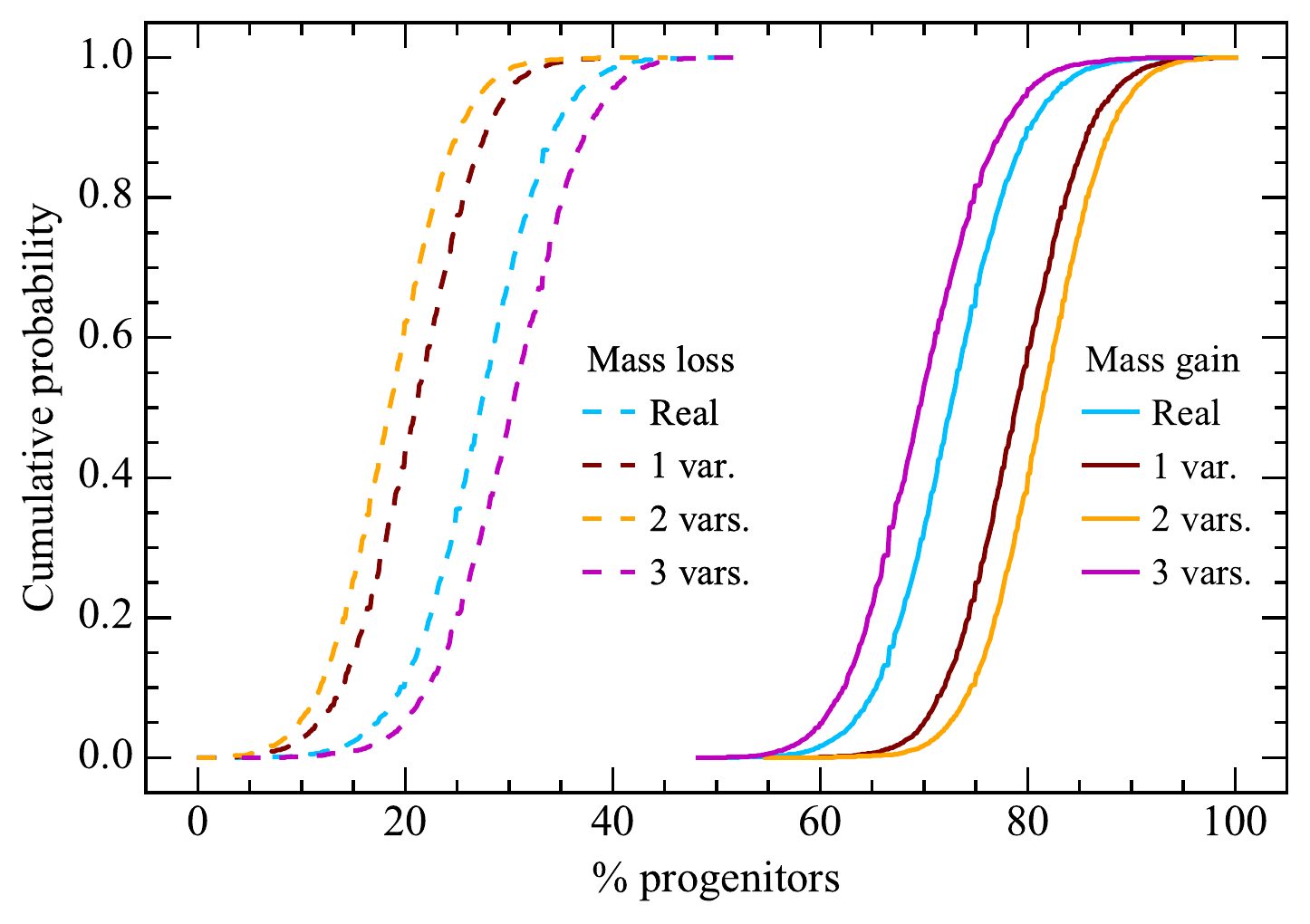}
  \includegraphics[width=\linewidth]{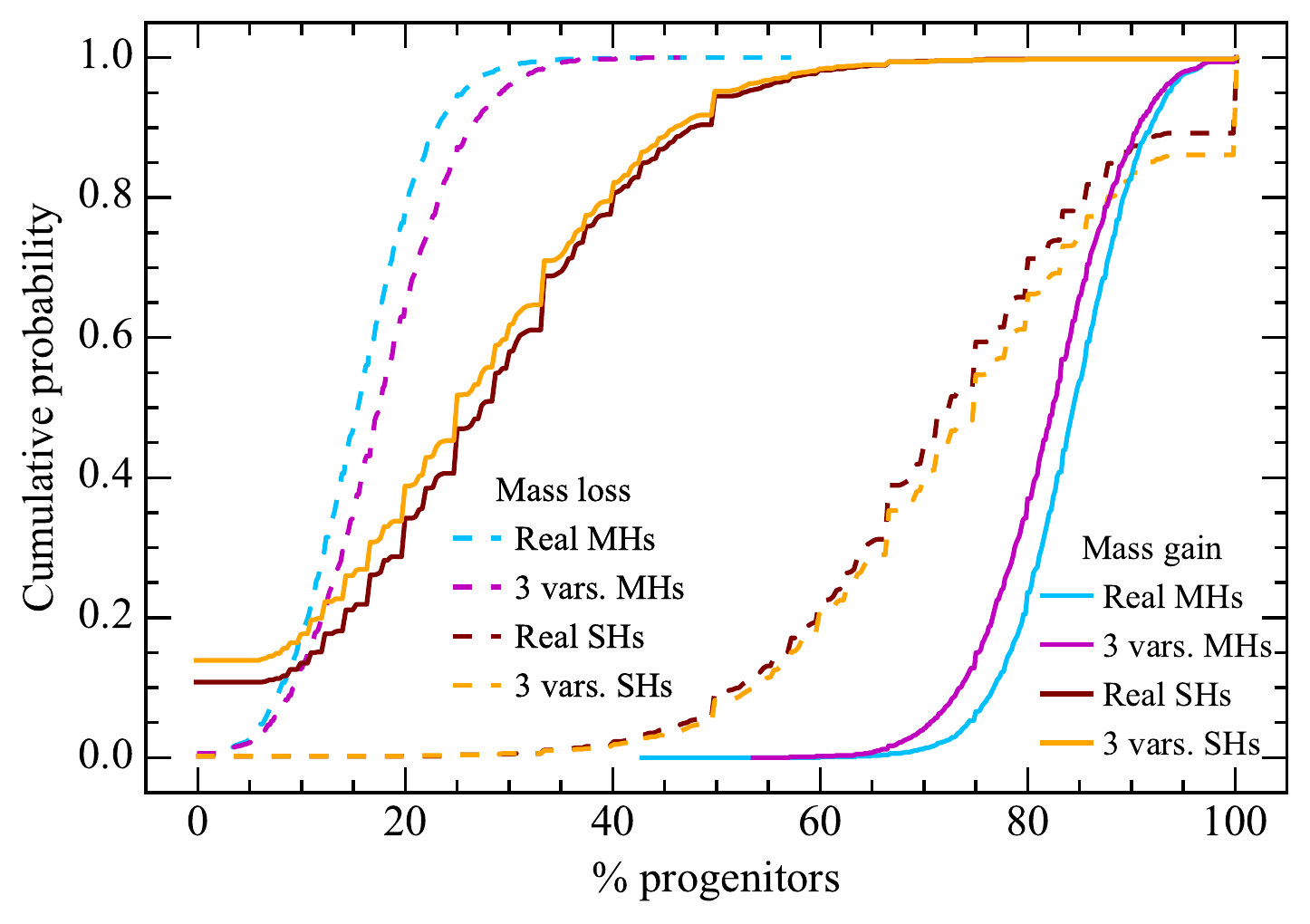}
  \caption{%
    Top: Cumulative probability of the mass gain and loss for all the progenitors in all the merger trees in the training dataset (real), and in the samples of generated merger trees with 1, 2, and 3~variables. 
    Bottom: Same for the progenitors that are main halos~(MH) and subhalos~(SH) in all the merger trees in the training dataset (real), and in the generated merger trees with three variables.
  }
  \label{fig:mass}
\end{figure}

We consider three different scenarios. 
The first configuration accounts for the mass reconstruction of halo progenitors, denoted henceforth as 1~variable. 
Then, we have trained our GAN with only the mass dataset. 
Next, we have introduced a second input, the distance to the main branch and trained the GAN with the mass and distance input dataset. 
This is the 2~variable configuration. 
Afterwards, the 3~variable configuration introduces the progenitor type as a new variable, and, once again, we trained our GAN with the dataset for the three aforementioned variables. 

Then, we evaluated the progenitor masses of the merger trees generated with 1, 2 and 3~variables according to Section~\ref{sec:evaluation}. 
We found that this variable is well reconstructed except for small numerical artifacts when there is no progenitor in a branch and the corresponding mass is not exactly zero, but a very small number. 

Next, we compared the distributions of the mass gain and loss for all the progenitors in the full sample of reconstructed merger trees for each of the above mentioned configurations with that of the training dataset (real trees). 
Note that the sample size of the generated merger trees is as large as that of the real trees. 
We represent the cumulative probability of these distributions in the top panel of Fig.~\ref{fig:mass}. 
The KS test (see first row of Table~\ref{tab:KStest}) shows that the 3~variable configurations yield the best mass reconstruction. 
To understand why the introduction of the progenitor type produce the best results, we have analyzed the mass gain and loss of the progenitors that are main halos (MH) and subhalos (SH) separately with the KS test. 
We show the cumulative probabilities of the MH and SH mass distributions in the bottom panel of Fig.~\ref{fig:mass}.  
According to the results of the KS test (second, MH, and third, SH, rows of Table~\ref{tab:KStest}), the 3~variables configuration gives the best fit to the trained data due to the GAN's ability to reproduce the mass gain and loss of SHs than those of the progenitors that are main halos. 

\begin{figure}[tb]
\centering
\includegraphics[width=\linewidth]{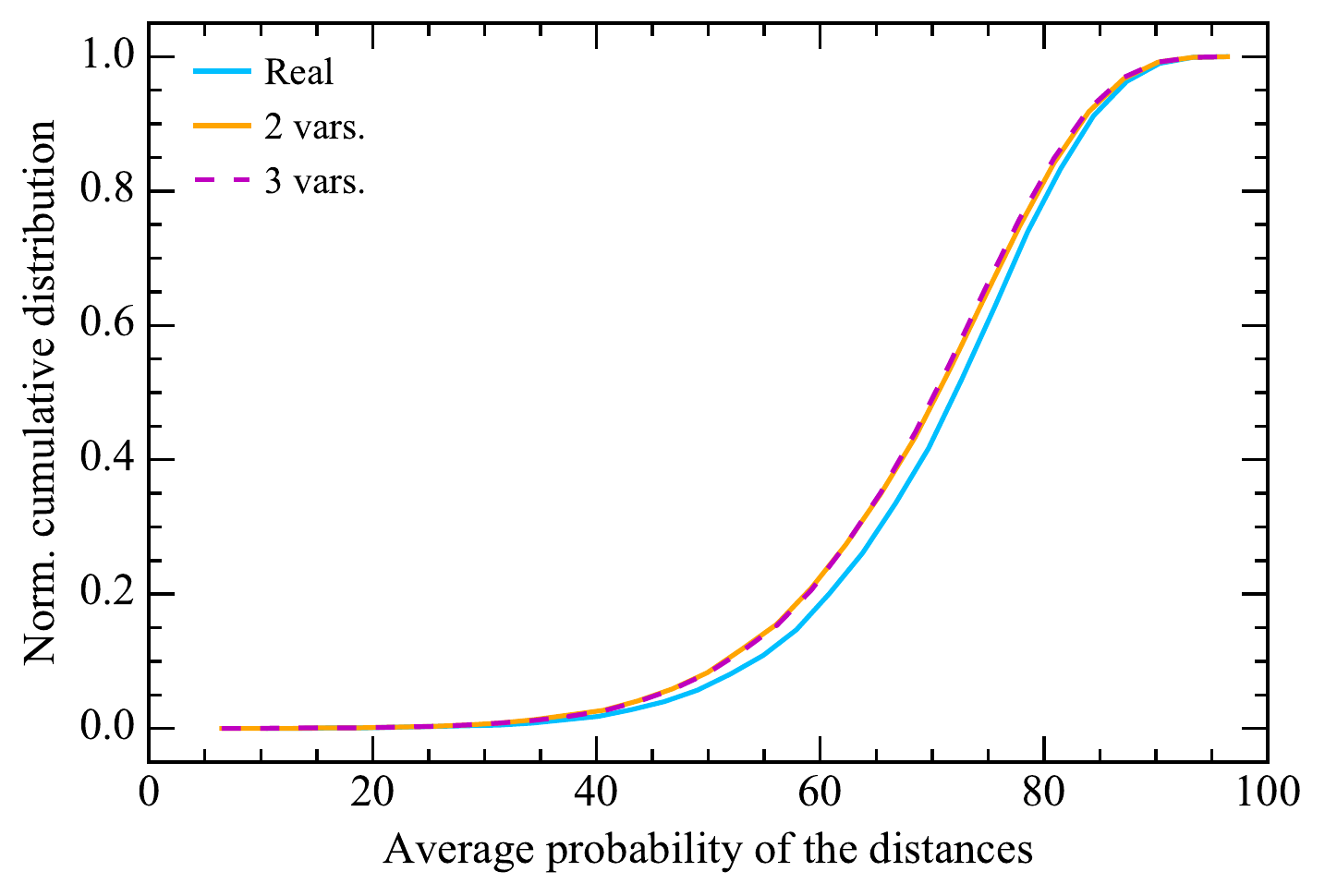}
\caption{Normalized cumulative distribution of the averaged probabilities of the distances for the real merger trees and those generated with 2 and 3~variables.}
 \label{fig:distance}
\end{figure}

To evaluate the reconstruction of the distance for the 2 and 3~variables configurations, as previously stated, we computed the probability distribution of the distance between merging progenitors for several snapshots before the merger takes place, for the training dataset and an equally large sample of reconstructed merger trees (see Appendix~\ref{sec:distance}). 
More specifically, we calculated the probability distribution (at all snapshots) of the distance to the main branch for the 
real sample, which will be our reference distribution in order to compute the probability of every distance in a merger tree of any sample. 
Then, for a given value of the distance, we calculated the average probability for each merger tree of the real, 2 and 3~variables samples. 
We show the normalized cumulative distribution of these average probabilities in Fig.~\ref{fig:distance}. 
We compared the real and fake distributions with the KS test, and show it on the fourth row of Table~\ref{tab:KStest}. 
Both generated distributions yield equally good results with the 2~variables configurations giving a slightly better goodness of the fit. 

\begin{figure}[tb]
\centering
\includegraphics[width=\linewidth]{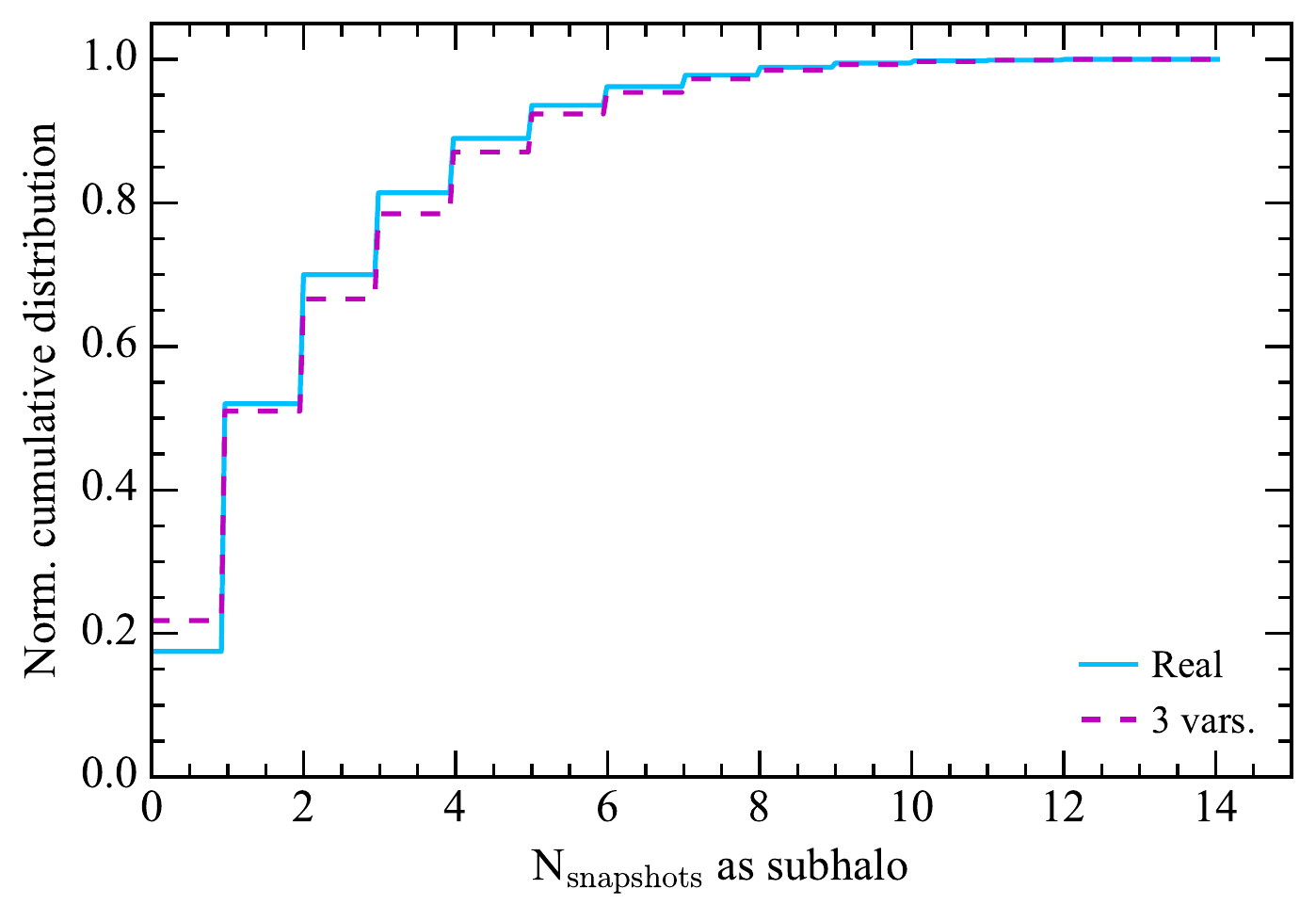}
\caption{Normalized cumulative distribution of the number of snapshots that a progenitor spend as a subhalo before the merger takes place.}
 \label{fig:subhalo}
\end{figure}

Finally, we can perform a last test on the ensemble of generated merger trees with 3 variables. This test focus only on the fair reproduction of the third variable, the progenitor type. We show in Fig.~\ref{fig:subhalo} the cumulative distribution of the number a snapshot a progenitor spend as subhalo before merging with another progenitor, for generated and real merger trees. 
Once again, we compared both distribution with the KS test (see last row of Table~\ref{tab:KStest}) and found a very good agreement. 

As final remark, note that there are unavailable tests in Table~\ref{tab:KStest} due to the requirement of additional variables which are out of the scope on the corresponding configuration.

\section{Conclusions}

We proposed a new framework for halo merger tree generation based on Generative Adversarial Networks that intends to learn the properties of more expensive simulations with a lower computational cost. 
Our proposal successfully generates the most important merger trees' properties, namely progenitor masses, distance of the progenitors to those in the main branch and the progenitor type, and inherits the best features of the trees of the EAGLE simulation suite. 
Despite restricting trees to have six branches in our experiments, our framework can straightforwardly be expanded to trees with a variable number of branches, provided that there is a large number of tree samples in the training dataset. 

A final test for our GAN generated merger trees would be to apply them to populate a semi-analytic model of galaxy formation and then  evaluate the quality of the resulting galaxies by comparing them with observations and/or hydrodynamically simulated galaxies. 
Nevertheless, given that the generated halo merger trees reproduce fairly the properties of the training dataset, specially merger trees generated with 3 variables, we expect as good outcomes as those obtained with trees from $N$-body simulations.

In addition, we have developed for the first time a set of criteria that takes into account up to three merger trees' properties  to compare different samples of trees. 
We applied these criteria to assess the quality of the learned representation of generated merger trees when compared to those in the training dataset and to measure the effect of training our GAN framework with additional merger trees' properties. 
It is worth remarking that this evaluation framework can also be used to compare merger trees of different large volume cosmological simulations and to assess the importance of the set of variables to take into account for a fair merger construction.

\section*{Acknowledgements}

SR acknowledges support from  MINECO/FEDER (Spain) under grant   AYA2015-63810-P  and from the Australian Research Council. 
SR also acknowledges funding from the European Union's Horizon~2020 Research and Innovation Programme
under the Marie Sklodowska-Curie grant agreement No.~734374 (LACEGAL-RISE) for a secondment at the  Pontificia Universidad Cat\'olica de Chile. 
JG acknowledges support from CONICYT project Basal~AFB-170002. 
This project has also received funding from the S\~{a}o Paulo Research Foundation (FAPESP) under grant No.~2016/19947-6, and from the Brazilian National Council for Scientific and Technological Development (CNPq) under grant No.~307425/2017-7.


\bibliographystyle{icml2019}

\bibliography{abrv,references}

\appendix
\counterwithin{figure}{section}
\counterwithin{table}{section}

\section{Details of the GAN Architecture}
\label{sec:architecture}

We selected a convolutional-based architecture, after having evaluated first a fully connected (FC) architecture with different number of layers. 
Even if the FC-based GAN successfully learned the merger tree structure, it failed to reproduce the correct mass range of the halos in branches different from the main branch. 

The discriminator was implemented with convolutional neural networks (CNNs) \cite{Krizhevsky2012} and the generator with deconvolutional layers. 
The layout of the encoder was also based on CNNs. 
We used a combination of column- and row-like filters in each neural network. 
These filters intend to reproduce the operations within a branch (column-wise filters), and between the branches to produce mergers (row-wise filters).
We present the parameters for the discriminator, encoder, and generator in Table~\ref{tab:params}. 
We tested different activation functions, where the Exponential Linear Unit (ELU) yielded the best results. 
The performance of the GAN was optimized when losses were calculated with cross entropy with logits and a batch of $100$ samples. 
It is worth remarking that the quality of the generated merger trees improved considerably with the introduction of an encoder. 

\begin{table}[t!]
\caption{Layout and parameters of the (top)~discriminator, (middle)~encoder, and (bottom)~generator networks, where $\nvar$ denotes the number of variables, $k$ is the kernel structure, and $s$ the number of strides.}
\label{tab:params}
\vspace{.15in}
\footnotesize
\begin{subtable}{\linewidth}
\centering
\begin{tabular}{lcl}
  \toprule
  Layer & Parameters & Output shape \\ 
  \midrule
  Input  &  & (29, 6, $\nvar$) \\
  Conv2D & $k$:(1, 3) \,$s$:1  &(29,6, 16) \\
  ELU &  &(29, 6, 16) \\
  Conv2D & $k$:(1, 3) \,$s$:1 & (29, 6, 32) \\
  ELU &  & (29, 6, 32) \\
  Conv2D & $k$:(3, 1) \,$s$:1 & (29, 6, 64) \\
  ELU &  & (29, 6, 64) \\
  Conv2D & $k$:(3, 1) \,$s$:1 & (29, 6, 128)\\
  ELU &  & (29, 6, 128) \\
  Conv2D & $k$:(3, 1) \,$s$:1 & (29, 6, 256)\\
  ELU &  & (29, 6, 256) \\
  Flatten & & (44544) \\ 
  FC &  & (1) \\
  \bottomrule
\end{tabular}
\vspace{15pt}
\label{tab:disc-params}
\end{subtable}
\begin{subtable}{\linewidth}
\centering
\begin{tabular}{lcl}
  \toprule
  Layer & Parameters & Output Shape \\ 
  \midrule
  Input  &  & (29, 6, $\nvar$) \\
  Conv2D & $k$:(1, 3) \,$s$:1  &(29, 6, 16) \\
  ELU &  &(29, 6, 16) \\
  Conv2D & $k$:(1, 3) \,$s$:1 & (29, 6, 32) \\
  ELU &  & (29, 6, 32) \\
  Conv2D & $k$:(3, 1) \,$s$:1 & (29, 6, 64) \\
  ELU &  & (29, 6, 64) \\
  Conv2D & $k$:(3, 1) \,$s$:1 & (29, 6, 128)\\
  ELU &  & (29, 6, 128) \\
  Conv2D & $k$:(3, 1) \,$s$:1 & (29, 6, 256)\\
  ELU &  & (29, 6, 256) \\
  Flatten & & (44544) \\ 
  FC &  & (100) \\
  \bottomrule
\end{tabular}
\vspace{15pt}
\label{tab:encoder-params}
\end{subtable}
\begin{subtable}{\linewidth}
\centering
\begin{tabular}{lcl}
  \toprule
  Layer & Parameters & Output Shape \\ 
  \midrule
  Input  &  & (100) \\
  FC &  & (44544)\\
  ELU &  & (44544) \\
  Deconv2D & $k$:(3, 1) \,$s$:1  &(29, 6, 128) \\
  ELU &  &(29, 6, 16) \\
  Deconv2D & $k$:(3, 1) \,$s$:1 & (29, 6, 64) \\
  ELU &  & (29, 6, 32) \\
  Deconv2D & $k$:(1, 3) \,$s$:1 & (29, 6, 32) \\
  ELU &  & (29, 6, 64) \\
  Deconv2D & $k$:(1, 3) \,$s$:1 & (29, 6, 16)\\
  ELU &  & (29, 6, 128) \\
  Deconv2D & $k$:(1, 3) \,$s$:1 & (29, 6, $\nvar$)\\
  Sigmoid &  & (29, 6, $\nvar$) \\
  \bottomrule
\end{tabular}
\label{tab:generator-params}
\end{subtable}
\end{table}

\section{Examples of Generated Merger Trees}
\label{sec:examples}

\begin{figure*}[tb]
  \centering
  \begin{subfigure}{.5\linewidth}
    \includegraphics[width=0.95\linewidth]{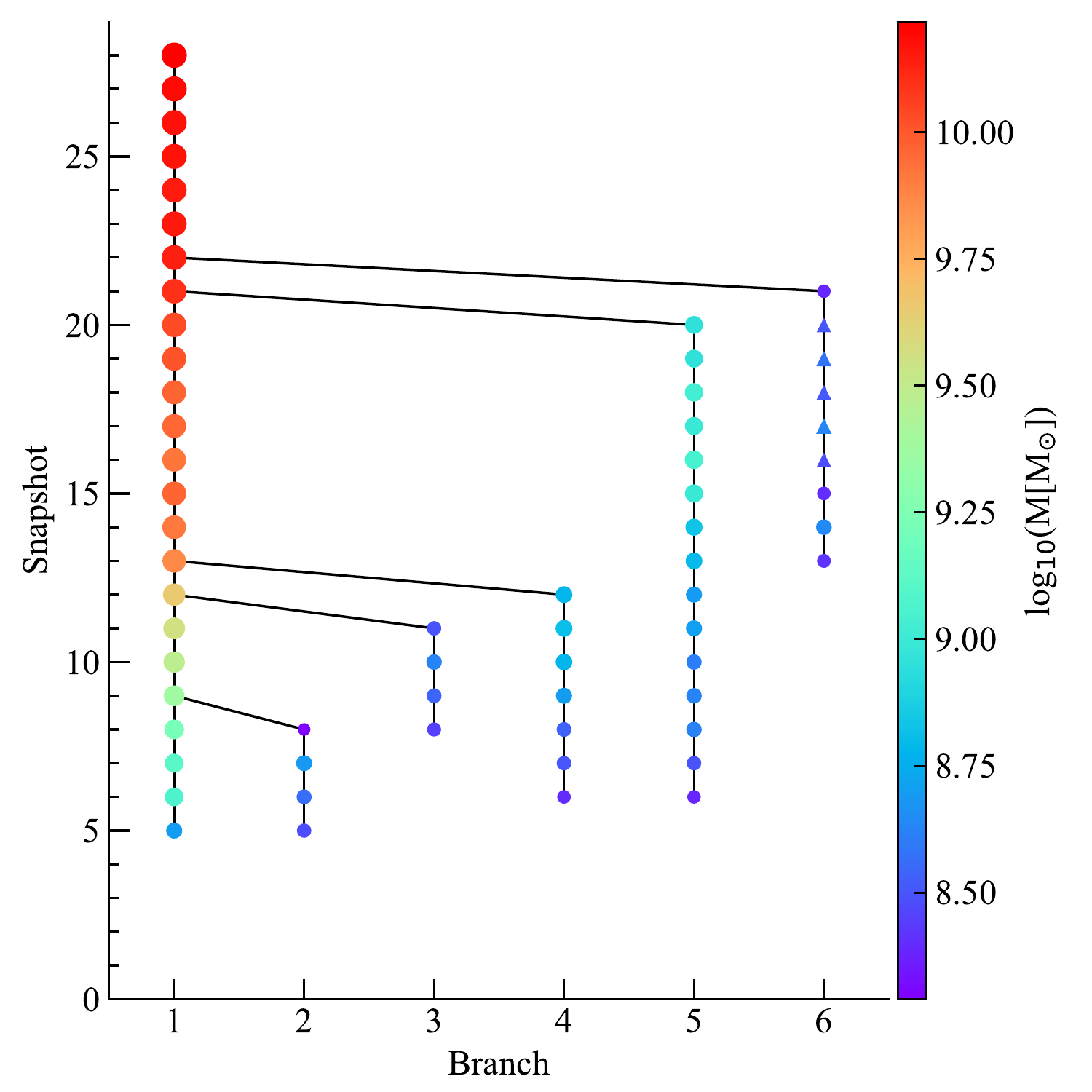}
  \end{subfigure}%
  \begin{subfigure}{.5\linewidth}
    \includegraphics[width=.95\linewidth]{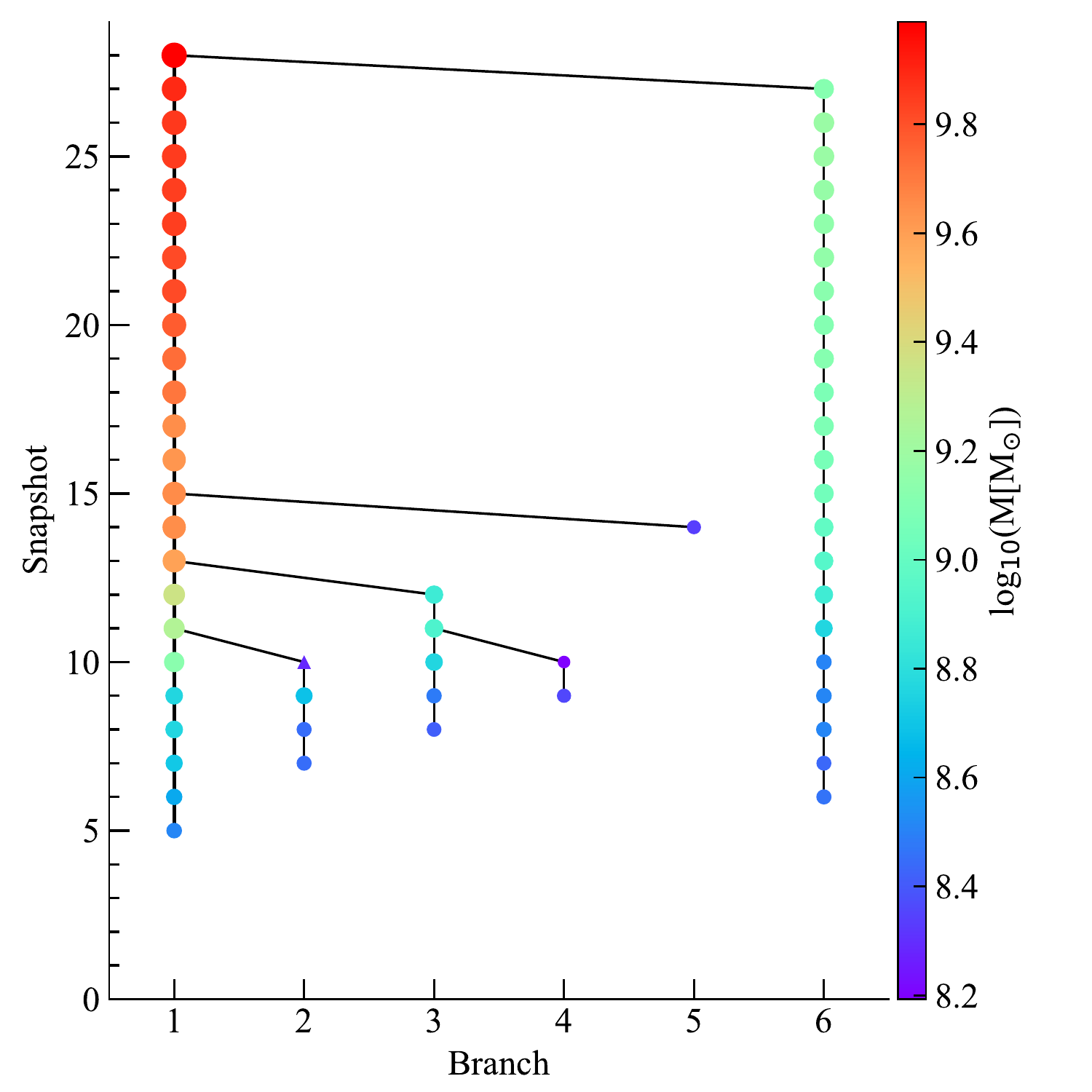}
  \end{subfigure}
  \begin{subfigure}{.5\linewidth}  
    \includegraphics[width=0.95\linewidth]{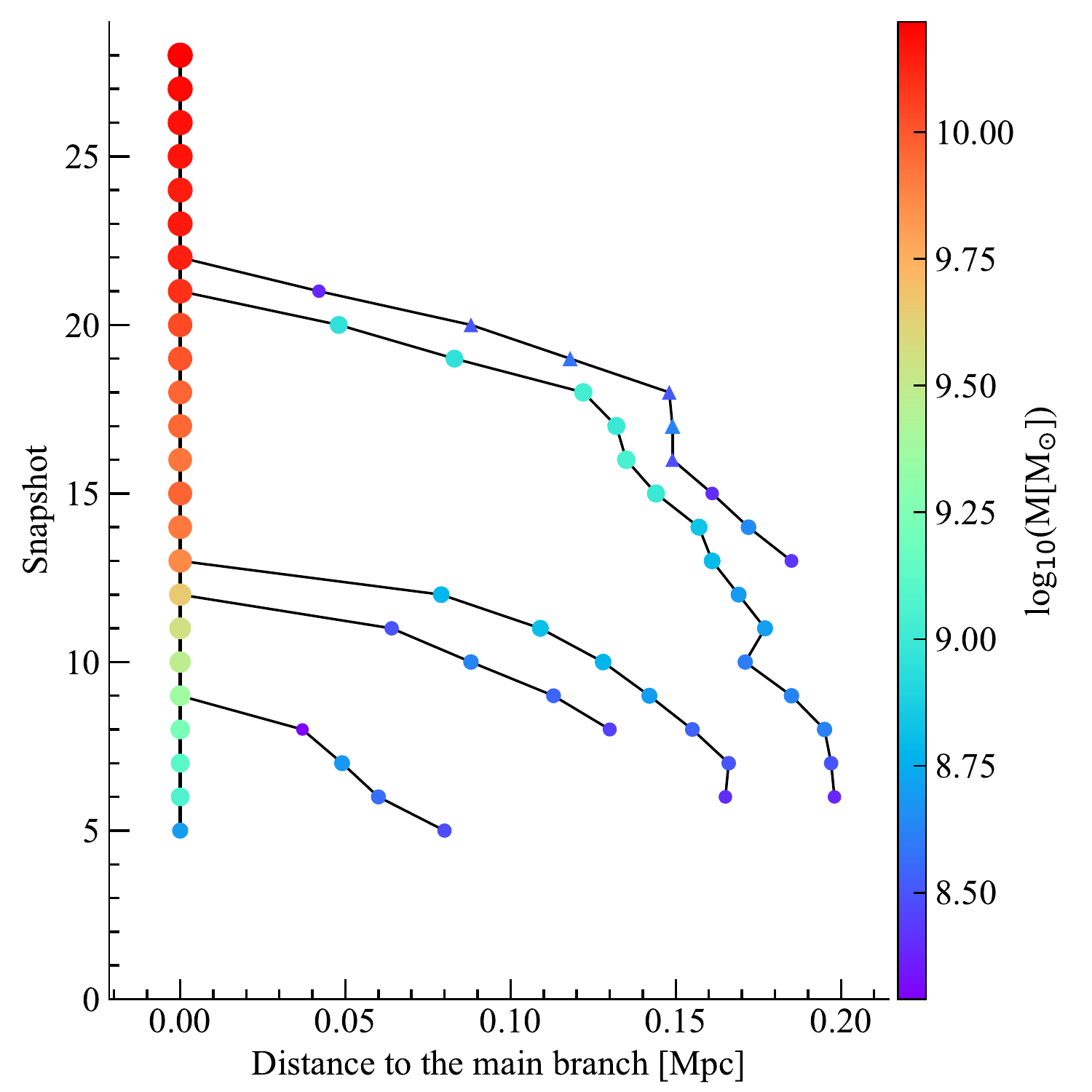}
  \end{subfigure}%
  \begin{subfigure}{.5\linewidth}  
    \includegraphics[width=0.95\linewidth]{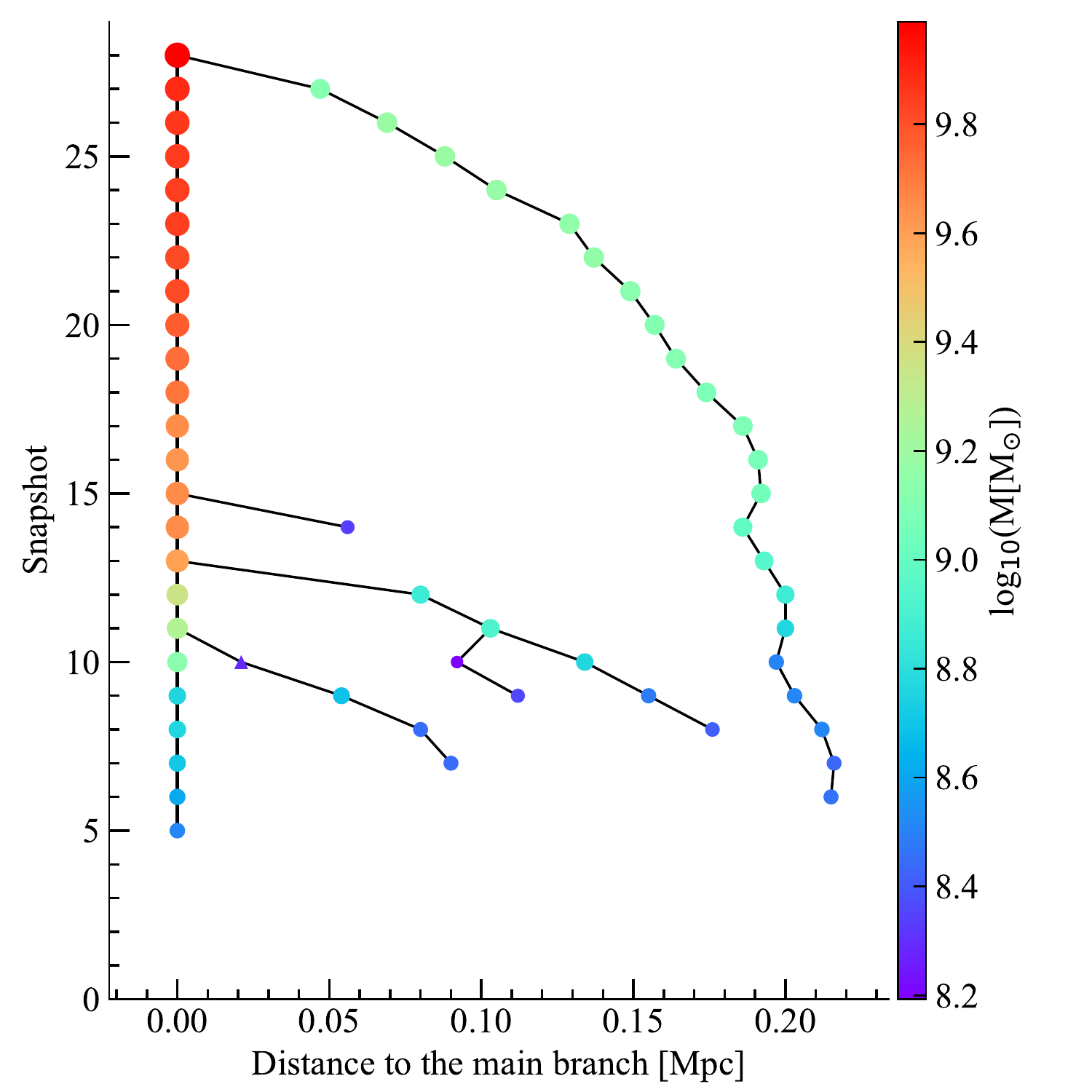}
  \end{subfigure}  
    \vspace{2pt}
  \caption{%
    Left: Merger tree generated with 3 variables, featuring no sub-branches, 
    Right: Another example of a merger tree generated with 3 variables, a sub-branch and a late merger characterize this tree. 
    The color map denotes progenitor masses and the progenitor type is represented by  circles (main halos) and triangles (subhalos). Top panels show merger trees in the plane snapshot vs.~branch, and bottom panels correspond to the same trees in the plane snapshot vs.~distance.
  } 
  \label{fig:mergertree2}

\end{figure*}

In Fig.~\ref{fig:mergertree2}, we show two additional examples of merger trees generated with three variables. 
The tree on the left panel has no sub-branches.
While the one on the right showcases a sub-branch in the third branch and a late merger (snapshot~$28$) of two large branches (main branch and branch~$6$). 
Note that the main branch hosts more massive progenitors. 

\section{Distributions of the Distance to the Main Branch}
\label{sec:distance}

\begin{figure}[tb]
\centering
\includegraphics[width=0.5\textwidth]{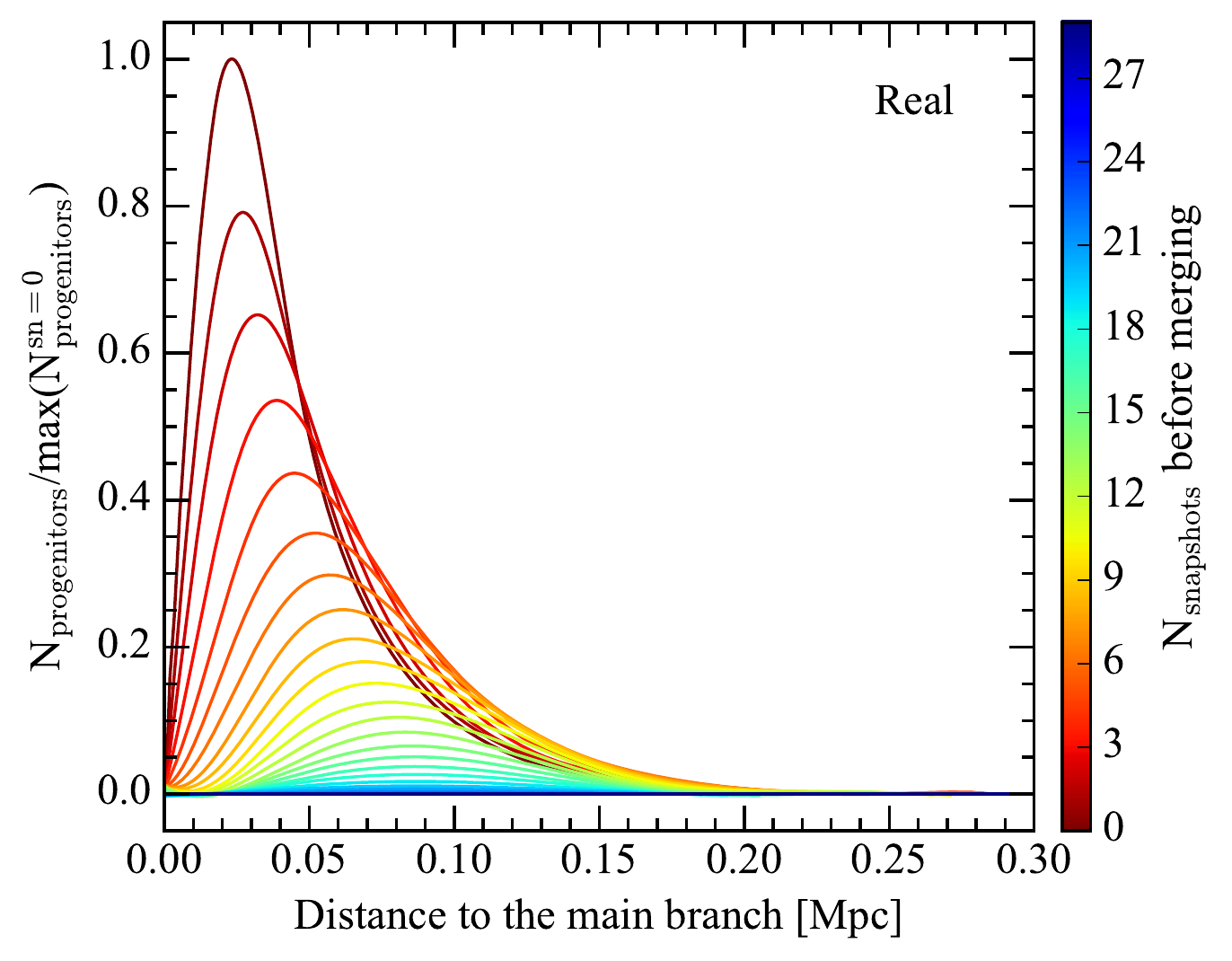}
\includegraphics[width=0.5\textwidth]{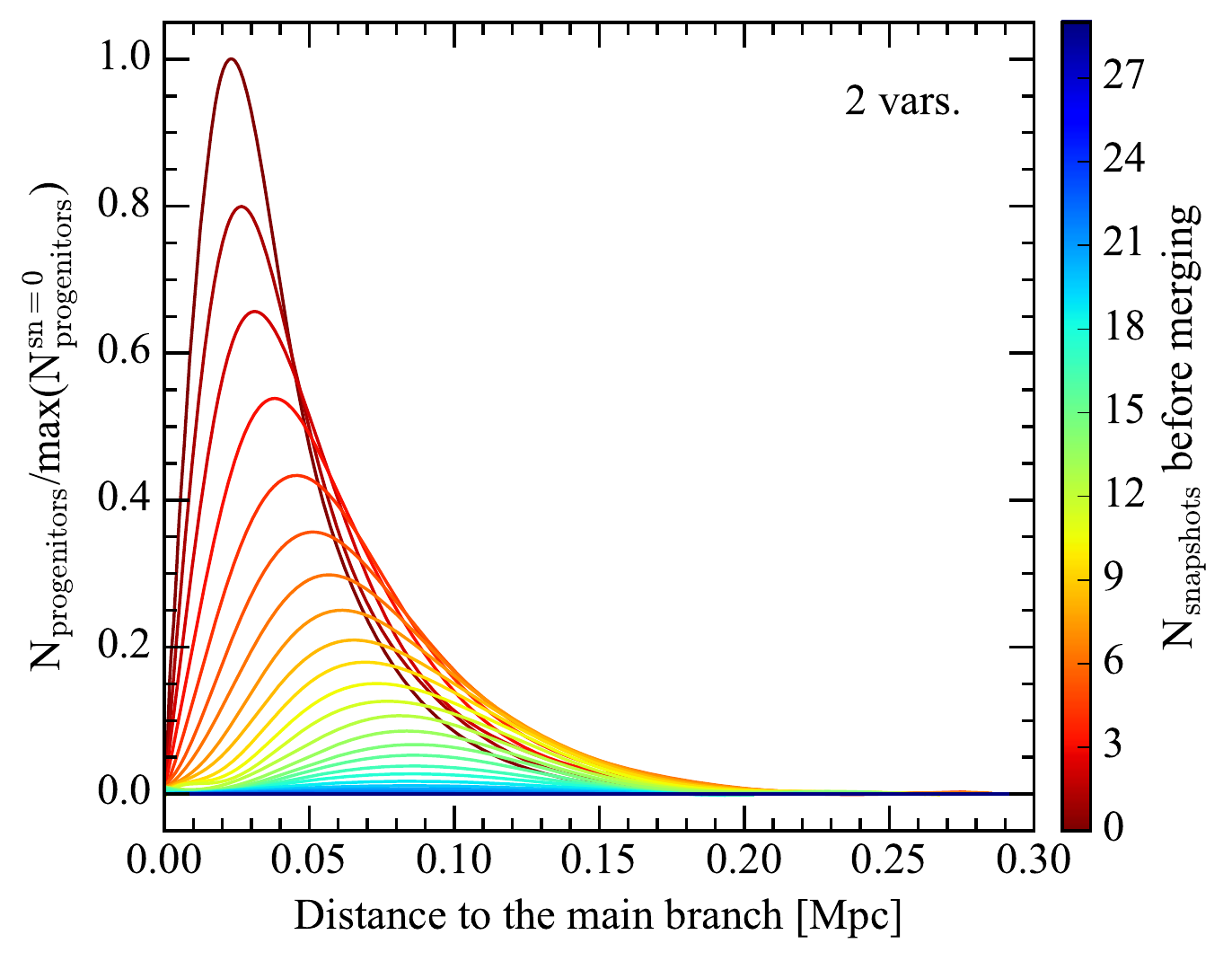}
\includegraphics[width=0.5\textwidth]{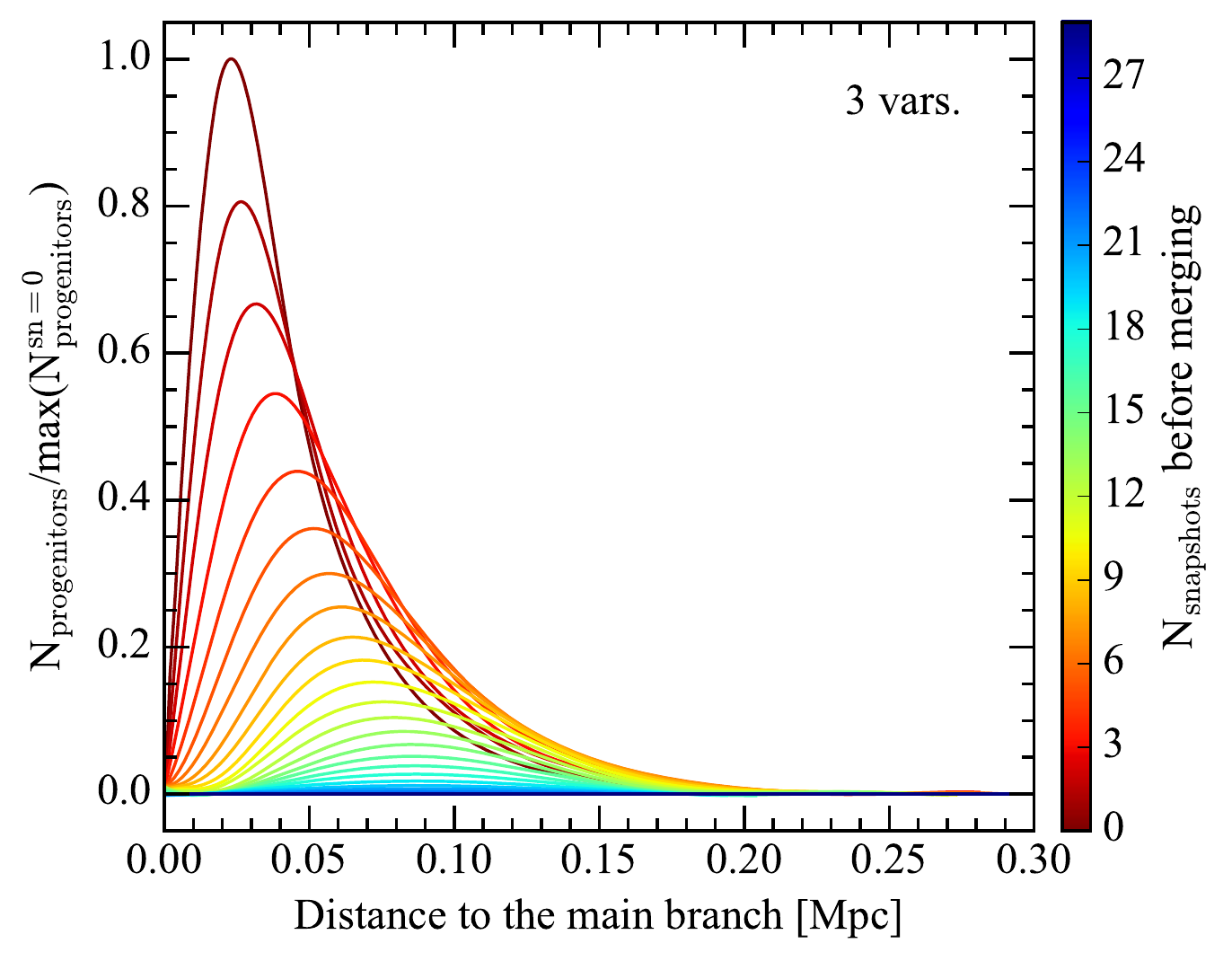}
\caption{%
  Distribution of the distance to the main branch for the real trees and those generated with 2 and 3~variables and several snapshots before the merger. 
  The number of progenitors is normalized by the maximum number of these halos at the snapshot~$0$. 
  The colormap denotes the number of snapshots before the merging event.
}
\label{fig:distancedistrib}
\end{figure}

In Fig.~\ref{fig:distancedistrib}, we show the distribution of the progenitor distance to the main branch for several snapshots before the merger takes place for real and  merger trees generated with 2 and 3~variables. 
The distribution is normalized by the maximum number of progenitors in the complete merger tree sample at the snapshot~$0$. 
Similar distributions but normalized by the number of progenitors in the merger tree sample at the timestep when the merger event occurs,  
were used to perform the KS test in Section~\ref{sec:results}.  
Note that the three distributions in Fig.~\ref{fig:distancedistrib} are very similar, the slight  differences among them are only noticeable with the KS test.

\end{document}